\documentclass[aps,prl,twocolumn,footinbib,superscriptaddress]{revtex4-1}
\usepackage[cp1251]{inputenc}
\usepackage{amsmath}
\usepackage{amssymb}
\usepackage{color,xcolor}
\usepackage{natbib}
\usepackage[colorlinks=true,bookmarks=false,citecolor=blue,urlcolor=blue]{hyperref} 
\usepackage{bm}
\usepackage{epsfig}
\usepackage{verbatim} 
\usepackage{morefloats}
\usepackage{graphicx}
\usepackage{bibentry}
\usepackage[mathscr]{euscript}
\usepackage{enumerate}

\usepackage{subfigure}

\usepackage{bm}

\def\dfrac{\displaystyle\frac}  

\renewcommand{\phi}{\varphi}
\newcommand{\addtuc}{School of Electrical and Computer Engineering, Technical University of Crete, Chania, Greece 73100}

\begin{document}

\title{Gradient catastrophe of nonlinear photonic valley-Hall edge pulses}

\author{\firstname{Daria A.} \surname{Smirnova}}
\affiliation{Nonlinear Physics Centre, Australian National University, Canberra ACT 2601, Australia}
\affiliation{Institute of Applied Physics, Russian Academy of Science, Nizhny Novgorod 603950, Russia}

\author{\firstname{Lev A.} \surname{Smirnov}}
\affiliation{Institute of Applied Physics, Russian Academy of Science, Nizhny Novgorod 603950, Russia}
\affiliation{Nizhny Novgorod State University, Gagarin Av. 23, Nizhny Novgorod,
603950 Russia}

\author{\firstname{Ekaterina O.} \surname{Smolina}}
\affiliation{Institute of Applied Physics, Russian Academy of Science, Nizhny Novgorod 603950, Russia}

\author{\firstname{Dimitris G.} \surname{Angelakis}}
\affiliation{Centre for Quantum Technologies, National University of Singapore, 3 Science Drive 2, Singapore 117543}
\affiliation{\addtuc}

\author{\firstname{Daniel} \surname{Leykam}}
\affiliation{Centre for Quantum Technologies, National University of Singapore, 3 Science Drive 2, Singapore 117543}

\begin{abstract}
We derive nonlinear wave equations describing the propagation of slowly-varying wavepackets formed by topological valley-Hall edge states. We show that edge pulses break up even in the absence of spatial dispersion due to nonlinear self-steepening. Self-steepening leads to the previously-unattended effect of a gradient catastrophe, which develops in a finite time determined by the ratio between the pulse's nonlinear frequency shift and the size of the topological band gap. Taking the weak spatial dispersion into account results then in the formation of stable edge quasi-solitons. Our findings are generic to systems governed by Dirac-like Hamiltonians and validated by numerical modeling of pulse propagation along a valley-Hall domain wall in staggered honeycomb waveguide lattices with Kerr nonlinearity.
\end{abstract}

\maketitle

The combination of topological band structures with mean field interactions not only gives rise to rich nonlinear wave physics~\cite{Smirnova2020APR,Saxena2020}, but is also anticipated to unlock advanced functionalities, such as magnet-free nonreciprocity~\cite{Chen2018}, tunable and robust waveguiding~\cite{Dobrykh2018,Shalaev2019}, and novel sources of classical and quantum light~\cite{Kruk2018,Smirnova2018,Mittal2018,Wang2019,Zeng2020,Lan2020}. Valley-Hall photonic crystals~\cite{Ma2016,Chen2017,Dong2017,Wu2017,Noh2018,Ni2018b,Chen2018b,Kang2018,Gao2018,Shalaev2018NatNano,He2018,Smirnova2020LiSA,Yang2020_Terahertz} show great promise for these applications, due to their ability to combine slow-light enhancement of nonlinear effects with topological protection against disorder, which limits the performance of conventional photonic crystal waveguides~\cite{Rechtsman2019,Sauer2020,Arregui2021}. 

Accurate modelling of pulse propagation through photonic crystal waveguides in the slow light regime requires taking into account the dispersion in the effective nonlinearity strength, which can induce effects such as pulse self-steepening and supercontinuum generation~\cite{Anderson1983,Panoiu2009,Travers2011,Husko2015}. Despite the proven importance of these effects in applications~\cite{Kivshar_book,Soljacic2004,RevModPhys.78.1135}, analysis of nonlinear light propagation in topological photonic structures most often assumes non-dispersive nonlinearities, in both the underlying material response~\cite{Plotnik2013,Kartashov2016,Leykam2016,Solnyshkov2017,Zhang2019,Tulop2020,Chaunsali2021,Guo2020,Xia2020,Mukherjee2020,Maczewsky2020} and effective models describing the propagation of nonlinear edge states~\cite{Ablowitz2013,Ablowitz2014,Lumer2016,Ivanov2020,Ivanov2020b}.
The simplest effective model is the cubic nonlinear Schr\"odinger equation (NLSE), which describes the self-focusing dynamics of edge wavepackets independent of the properties of the topological band gap, such as its size.

More sophisticated effective models such as nonlinear Dirac models explicitly include the nontrivial spin-like degrees of freedom required to create topological band gaps~\cite{Maravero2016,Bomantara2017,Sakaguchi2018,RingSoliton2018,Smirnova2019LPR}. In the bulk, the nonlinear Dirac model supports  self-induced domain walls and solitons whose stability and degree of localization are sensitive to the gap size. Infinitely-extended (plane wave-like) nonlinear edge states can be obtained analytically and exhibit similar features. However, the nonlinear dynamics of localized edge pulses within the nonlinear Dirac model framework were not yet considered.

In this paper we study an analytically-solvable nonlinear Dirac model (NDM) describing topological edge pulses, revealing that nonlinear topological edge states' exhibit a self-steepening nonlinearity when the pulse self-frequency modulation becomes comparable to the width of the topological band gap. The self-steepening nonlinearity leads to the formation of a gradient catastrophe of edge wavepackets within a finite propagation time proportional to the pulse width. Taking the weak spatial dispersion of the topological edge modes into account regularizes the catastrophe and results in the formation of stable edge solitons for sufficiently long pulses. We validate our analysis using numerical simulations of beam propagation in a laser-written valley-Hall waveguide lattice, demonstrating that this effect should be observable even for relatively weak nonlinearities. Our findings suggest valley-Hall photonic crystal waveguides will provide a fertile setting for observing and harnessing higher-order nonlinear optical effects.

We consider a generic continuum Dirac model of topological photonic lattices
with incorporated nonlinear terms.
The evolution of a spinor wavefunction ${\bf{\Psi}}(x,y,t) = [\Psi_1(x,y,t); \Psi_2(x,y,t) ]^T$ 
in the vicinity of a band crossing point (topological phase transition) is 
governed by the nonlinear Dirac equation~\cite{Maravero2016,RingSoliton2018,Smirnova2019LPR}
\begin{align} 
\label{eq:Dirac_system}
&i \partial_{t} {\bf{\Psi}} = \left( \hat{H}_{D}(\delta {\bf{k}}) + \hat{H}_{N\!L} \right) {\bf{\Psi}} ;\\
\label{eq:Dirac_Hamiltonian}
&\hat{H}_{D}(\delta {\bf{k}}) =\delta k_{x} \hat{\sigma}_{x}+\delta k_{y} \hat{\sigma}_{y}+M \hat{\sigma}_{z},
\end{align}
where $\delta {\bf{k}}=\left(\delta k_{x}, \delta k_{y}\right) \equiv-i\left(\partial_{x}, \partial_{y}\right)$ is the momentum, $M$ is a detuning between two sublattices or spin states, and $\hat{H}_{N\!L}=-g \, \mathrm{diag}[\left|\Psi_1 \right|^2;\left|\Psi_2 \right|^2 ]$ is a local non-dispersive Kerr nonlinearity.

\begin{figure}[b]
{\includegraphics[width=1\columnwidth]{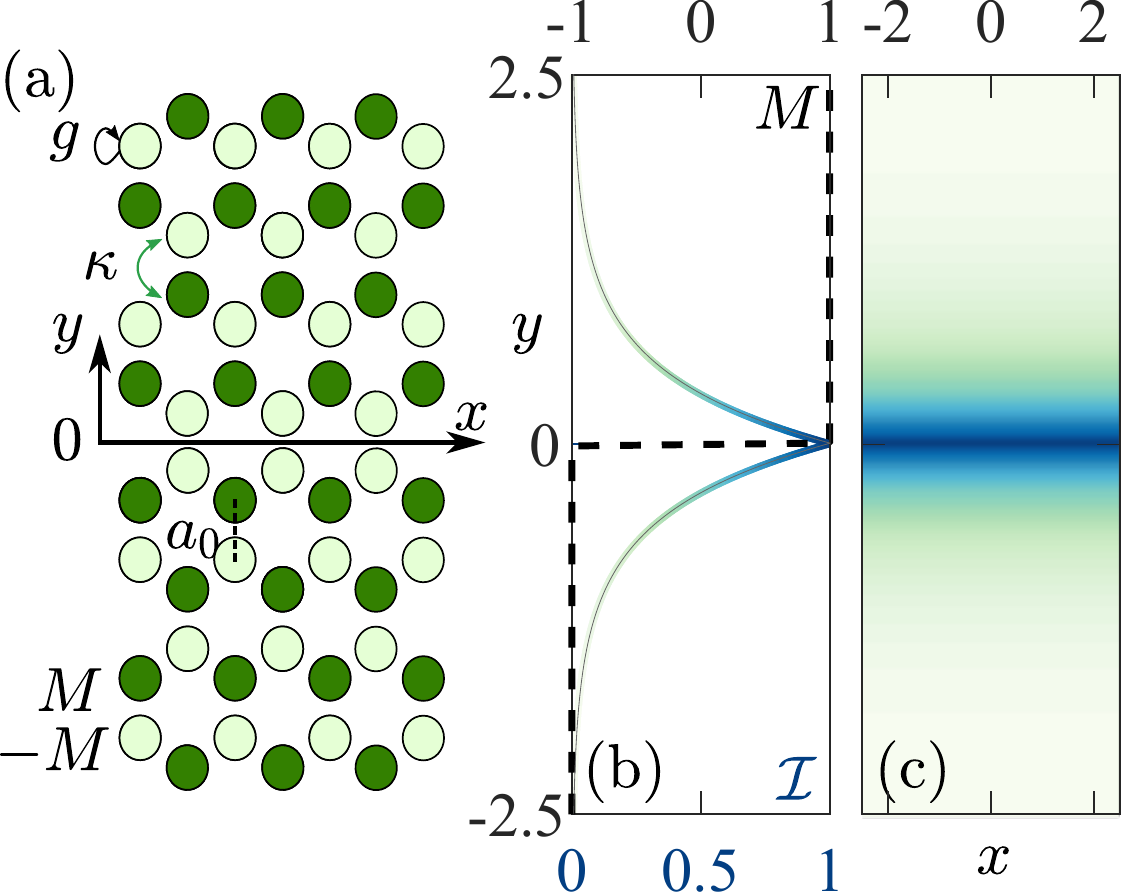}} 
\caption{(a) Dimerized photonic graphene lattice with staggered sublattice potential.
Here $\kappa$ denotes the tunneling coefficient between waveguides, $a_0$ is the distance between neighboring waveguides, and the round arrow $g$ illustrates the nonlinear self-action effect.
(b) Transverse profile $\mathcal{I}(y)$ and (c) in-plane intensity distribution $\mathcal{I}(x,y)=|\psi_1(x,y)|^2+|\psi_2(x,y)|^2$ of the nonlinear edge wave propagating along the $x$ axis and 
bound to the domain wall located at $y = 0$, where 
the effective mass $M(y)$ changes sign.
Parameters are $M_0 = 1, ~ g =0.75 $. }
\label{fig:graphene}
\end{figure}

As a specific example, 
the model~\eqref{eq:Dirac_system} 
can be implemented in nonlinear photonic graphene with a staggered sublattice potential as illustrated in Fig.~\ref{fig:graphene}(a). 
A dimerized graphene lattice 
is composed of single-mode dielectric waveguides with local Kerr nonlinearity.
The effective mass $M$ characterises a detuning between propagation constants in the waveguides of two sublattices.
The form
of Hamiltonian operator~\eqref{eq:Dirac_Hamiltonian} assumes
normalization of the transverse coordinates $x,y$ and evolution variable 
in the propagation direction $t\sim z/v_D$ to the Dirac velocity $v_D = 3 \kappa a_0 /2$ defined by the lattice parameters, 
a coupling constant $\kappa$ and a distance $a_0$ 
between two neighboring waveguides~\cite{Smirnova2019LPR,Smirnova2020LiSA}. This continuum model is valid provided $|\kappa| \gtrsim 2|M|$~\cite{Smirnova2020LiSA}. 

The {\it valley-Hall domain wall} 
is created 
between two insulators characterized by different signs of the mass. We take $M(y>0) = M_0$ and $M(y<0) = -M_0$ and without loss of generality assume $M_0>0$ in the upper half-space. 
In Ref.~\cite{Smirnova2019LPR}  we derived exact analytic solutions for the propagating nonlinear valley edge modes bound to the interface $y=0$: ${\bf{\Psi}}(y,x,t) = [\psi_1^0(y,\omega,k); \psi_2^0(y,\omega,k) ]^T e^{i k x - i \omega t}$. 
Based on the close connections between nonlinear edge states and self-trapped Dirac solitons in topological band gaps revealed in Ref.~\cite{Smirnova2019LPR}, the nonlinear edge mode dispersion can be obtained as
\begin{equation} \label{eq:nonl_dispn}
\omega + k = - g \mathcal{I}_1/2
\end{equation}
with the transverse profile of the edge state determined by the intensity at the interface $\mathcal{I}_1 = |\psi_{1,2}(y=0)|^2$ (see  Supplemental Material~\cite{Suppl}). In Figs.~\ref{fig:graphene}(b,c) we show 
the plane wave-like profile of the nonlinear edge mode with fixed wavenumber $k$ parallel to the edge.

Using the global parity symmetry with respect to the interface and analytical solutions for the edge states~\cite{Suppl}, we calculate two characteristics of the edge states via integration in the upper half plane $y>0$: the power $\mathcal{P}$ and spin ${{S}}_x$,
\begin{equation}
{\mathcal P} = \int_{0}^{\infty}  {\boldsymbol{\psi}}^\dagger {\boldsymbol{\psi}}dy, \quad {{S}}_x  = \frac{1}{2} \int_{0}^{\infty} { {\boldsymbol{\psi}}^\dagger  \hat{\sigma}_x {\boldsymbol{\psi}} } dy\:,
\end{equation}
and identify a functional relation $ {{S}}_x  \left({\mathcal P} \right)$ between them: 
\begin{equation} \label{eq:Sx_on_P}
 {{S}}_x = -  \frac{1}{g} \arcsin \left[ \frac{1}{\sqrt{2}} \sin\left({\mathcal P} \frac{g}{\sqrt{2}}\right) \right]\:.
 \end{equation}
Crucially, this relation is independent of the wavevector $k$, which allows us to develop a slowly varying envelope approximation to describe the nonlinear dynamics of finite edge wavepackets. 
Using Eq.~\eqref{eq:Dirac_system}, it can be shown that the integral characteristics obey the following evolution equation:
\begin{align} 
\label{eq:evol}
{\partial}_t \mathcal{P}  &= - 2 {\partial}_x {S_x} (\mathcal{P}).
\end{align}  
Next, we assume $\mathcal{P}(x,t)$ and $S_x(x,t)$ are slowly varying functions of the local frequency and wavenumber, such that Eq.~\eqref{eq:Sx_on_P} remains valid to a first approximation for smooth $x$-dependent field envelopes. 
Plugging Eq.~\eqref{eq:Sx_on_P} into Eq.~\eqref{eq:evol}, and using Eq.~\eqref{eq:nonl_dispn} assuming weak nonlinearity $g \mathcal{I}_1 \ll M_0$, we obtain the simple nonlinear wave equation for the longitudinal intensity profile $\mathcal{I}_1(x,t)$: 
 \begin{equation}
 \label{eq:sim_waves_for_I_1}
\partial_t \mathcal{I}_1-\partial_x  \mathcal{I}_1 \left(1-g^2 \mathcal{I}_1^2/\left(4 M_0^2\right)\right)=0. 
\end{equation}
Equation~\eqref{eq:sim_waves_for_I_1} suggests that the evolution of finite wavepackets propagating along the $x$ axis 
is accompanied by steepening of the trailing wavefront up to the development of {\it a gradient catastrophe}. 
Note, in the linear case $g=0$, Eq.~\eqref{eq:sim_waves_for_I_1} shows that the edge wavepacket of any shape does not diffract and propagates along the domain wall with constant group velocity $v = -1$,
being granted with topological 
robustness. Alternatively, Eq.~\eqref{eq:sim_waves_for_I_1} can be derived using asymptotic methods based on a series expansion of the spinor wavefunction~\cite{Suppl} 
\begin{equation} \label{eq:Psi12_series}
\Psi_{1,2} (x,y,t)  =  \pm    a (\xi; \tau_n )  e ^{-M_0 |y|}+ \sum_{n=1}^{\infty} \mu^n \Psi_{1,2}^{(n)}  \left( y; \xi; \tau_n  \right)\:, 
\end{equation}
where we have introduced a small parameter $\mu \sim {g \mathcal{I}_1}/{M_0}$, a hierarchy of time scales: $\tau_{n} = \mu^n t$, and assumed a smooth dependence 
of the spinor components on $t$ in the moving coordinate frame $(\xi \equiv x + t, y)$. 

\begin{figure}[b!]
\centering{
\includegraphics[width=1\linewidth] {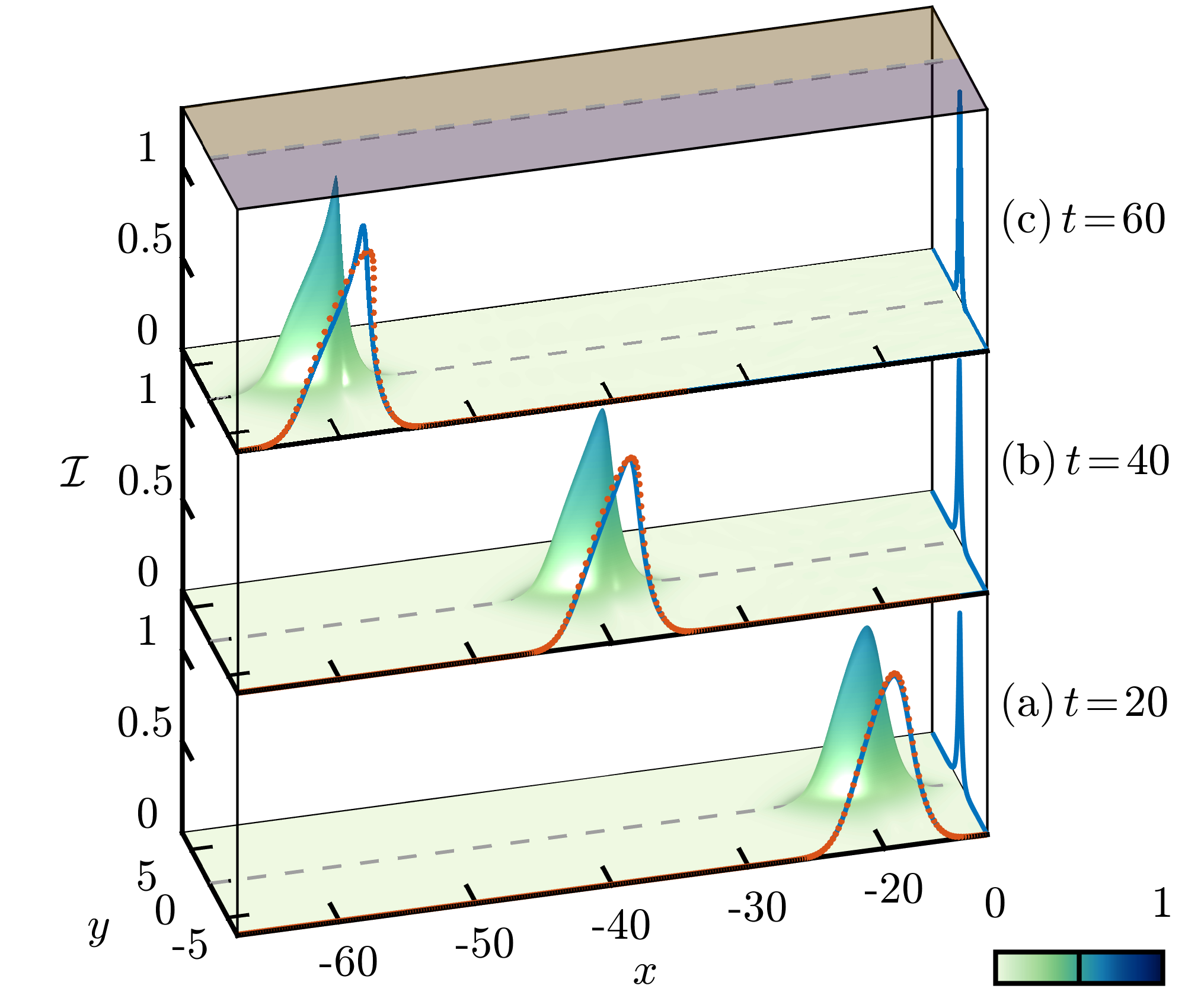} 
}
\caption{Gradient catastrophe development. 
The Gaussian pulse with $F_0=1$, $\Lambda_0=3.5$ is launched at $t=0$. 
Slices color code intensity distributions $\mathcal{I}(x,y)$ at the 
given moments: (a) $t=20$, (b) $t=40$, (c) $t=60$. Cuts along the domain wall at $y=0$ show consistency of the numerical solution (blue curves) with the solution of the nonlinear simple wave equation~\eqref{eq:sim_waves_for_I_1} for the intensity (red dotted lines). 
Parameters are $M_0=1$, $g=0.75$. In Figs.~\ref{fig:fig_2},~\ref{fig:fig_3}, dashed lines 
trace the domain wall separating spatial domains with 
masses of the opposite sign as indicated by shading with different colors on the top surface.  
}
\label{fig:fig_2}
\end{figure}

To illustrate the key effect of the gradient catastrophe 
captured by Eq.~\eqref{eq:sim_waves_for_I_1}, we model the time dynamics of an edge wavepacket using a custom numerical code, applying a split-step scheme and 
the fast Fourier transform to solve Eq.~\eqref{eq:Dirac_system}. Figure~\ref{fig:fig_2} shows evolution of the initial distribution set in the form of the 
edge state 
across the interface 
with the Gaussian envelope $\mathcal{I}_1^0(x,t=0) = F_0 e^{- x^2/\Lambda_0^2}$ along the $x$ axis: $\boldsymbol{\psi}(y,x,t=0) = \left[\psi_1^0\left(y,\omega = - {g \mathcal{I}^0_1 \left(x\right)}/{2} ,0\right); \psi_2^0\left(y,\omega = - {g \mathcal{I}^0_1 \left(x\right)}/{2},0\right) \right]^T$. Plugging the Gaussian distribution into Eq.~\eqref{eq:sim_waves_for_I_1}, we may estimate the pulse breakdown time $t_{*}$ analytically:
\begin{equation} \label{eq:t*}
t_{*} = 2 \sqrt{e} \Lambda_0 \left( \dfrac{M_0} {g F_0}\right)^2.
\end{equation}
Thus, pulse breakdown occurs for finite wavepackets when the peak nonlinear frequency shift becomes comparable to the size of the topological band gap. As the pulse propagates its tail becomes increasingly steep, developing a discontinuity (i.e. a shock) in a finite time. The numerical solution of Eq.~\eqref{eq:Dirac_system} is fully consistent with our analytical considerations, see Fig.~\ref{fig:fig_2}.

Weak spatial dispersion effects 
serve as a possible mechanism regularizing the gradient catastrophe, resulting in the formation of solitons.
For honeycomb photonic lattices, 
dispersion is accounted for by introducing off-diagonal 
second-order derivatives with the coefficient $\eta = (6 \kappa)^{-1}$ into the Dirac model~\eqref{eq:Dirac_system}:
\begin{equation} 
\hat{H}_{\text{disp}}\! = \left( \!{\begin{array}{*{22}{c}}
{0} & { - \eta \left( - i \partial_x +  \partial_y \right)^2}\\
{- \eta \left( i \partial_x + \partial_y \right)^2}  & {0}
\end{array}} \! \right).
\end{equation}
Assuming $\eta M_0 \sim \mu^2$ and developing a perturbation theory with expansion~\eqref{eq:Psi12_series}, we derive an evolution equation governing the complex-valued amplitude $a(\xi,t)$: 
\begin{equation}
\label{eq:solw}
i\left({\partial_{t} a} +\dfrac{g^2 |a|^2}{32 M_0^2}  {  a {\partial_{\xi}} |a|^2}  \right)  + \dfrac{g}{4} |a|^2 a  + \eta \left({\partial_{\xi^2}} a - M_0^2 a\right) = 0
\:,
\end{equation} 
which differs from the conventional cubic nonlinear Schr\"odinger equation by the second higher-order nonlinear term responsible for the phase modulation and self-steepening. 
This equation enables analysis of both the modulational instability of nonlinear plane-wave-like edge states, and the formation of edge quasi-solitons~\cite{Suppl}. 

\begin{figure}[b!]
\centering{
\includegraphics[width=1\linewidth] {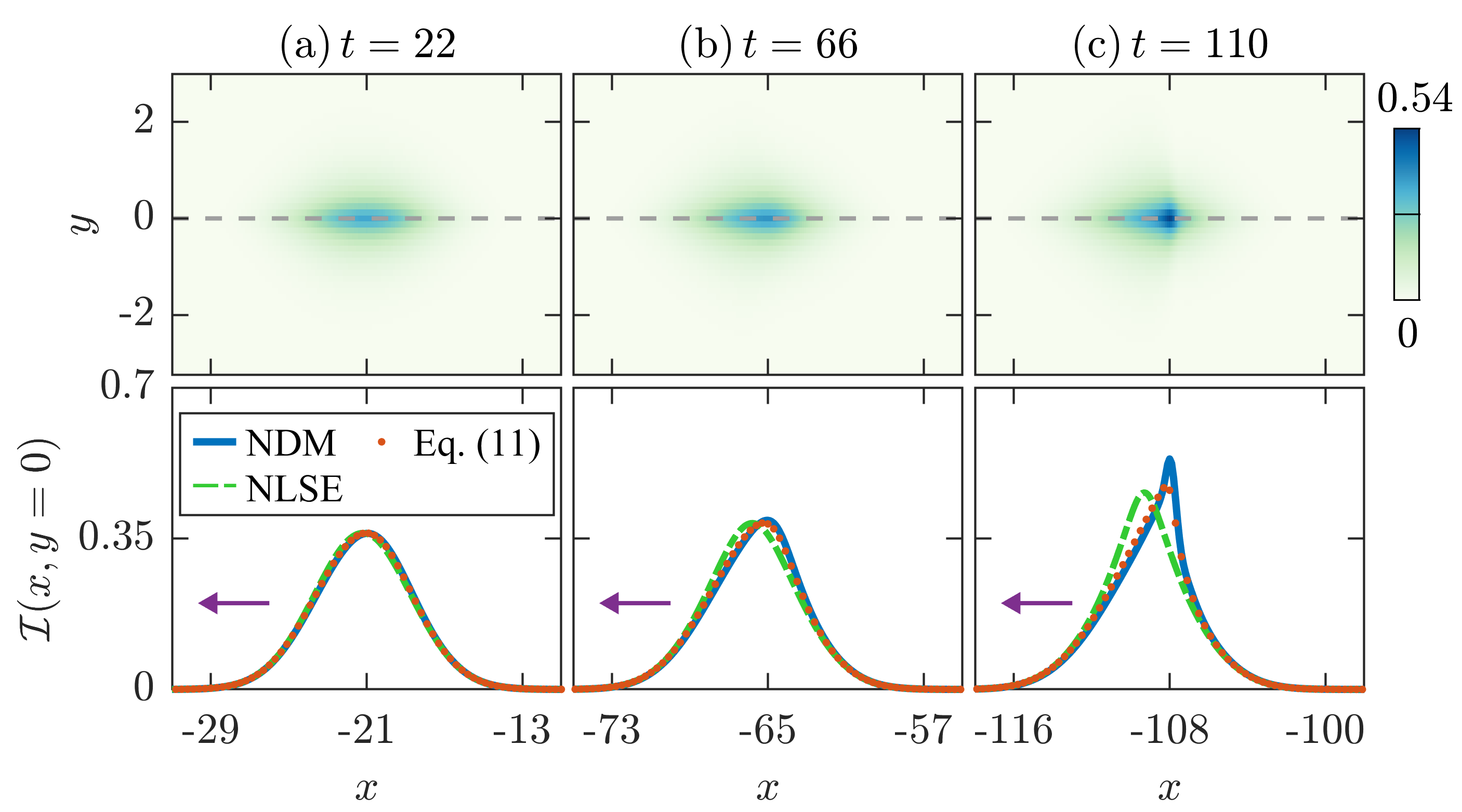} 
}
\caption{Nonlinear pulse transformation in the domain-wall problem with dispersion. Snapshots show intensity distributions in-plane $\mathcal{I}(x,y)$ (top row) and along the domain wall $\mathcal{I}(x,y=0)$ (bottom row) at the given times: (a) $t=22$, (b) $t=66$, (c) $t=110$. Overlaid curves are the pulse envelopes calculated by using NLSE (green dashed) and 
Eq.~\eqref{eq:solw} (red dotted). The Gaussian pulse with $F_0 = 0.18$, $\Lambda_0 = 5 /\sqrt{2}$ is launched at $t=0$. Parameters are $M_0=1$, $g=1$, $\eta = 0.001$.}
\label{fig:fig_3}
\end{figure}

To verify the validity of the modified NLSE~\eqref{eq:solw} we consider the propagation of a Gaussian pulse in Fig.~\ref{fig:fig_3}. The conventional NLSE, which lacks the self-steepening term, only exhibits self-focusing and gradual self-compression of the pulse. On  the other hand, the modified Eq.~\eqref{eq:solw} correctly reproduces the growing asymmetry of the edge pulse as it propagates. We show in the Supplemental Material how the self-steepening leads to the break-up of wide pulses, resulting in the radiation of part of its energy into bulk modes, with the remainder forming a quasi-soliton which continues to propagate along the edge and is capable of traversing sharp bends. We note that even after the initial pulse breakup, self-steepening terms can influence the soliton stability and soliton-soliton interactions~\cite{Kivshar_book}. 

\begin{figure}[t!] 
\centering{
\includegraphics[width=1\linewidth] {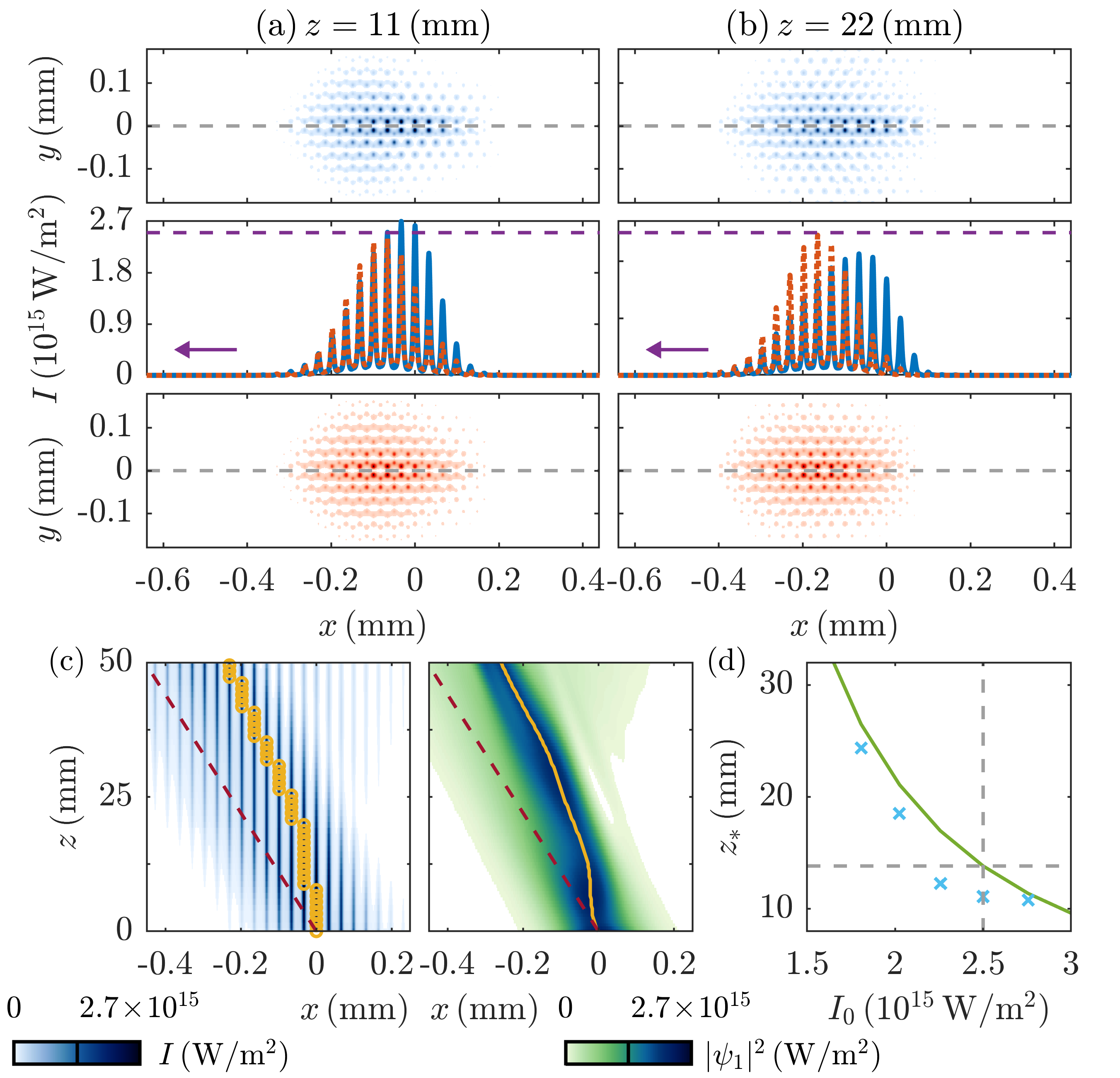}  
}
\caption{Nonlinear dynamics of the optical beam at the valley-Hall domain wall at a zigzag interface in a honeycomb lattice of laser-written waveguides. 
The input beam has Gaussian envelope along the domain wall with maximum intensity $I_0 = 2.5 \times 10^{15}$~W/m$^2$. 
(a,b) Intensity distributions of the nonlinear beam  
(top panels, blue line) at propagation distances (a) $z=11$~mm (left column) and (b) $z=22$~mm (right column). For comparison, the linear beam evolution, i.e. with nonlinearity switched off, is shown on bottom panels and with dotted red lines. Purple arrows points to the direction of motion. Dashed grey line depicts the domain wall. (c) Intensity distribution at the interface obtained in modeling of the paraxial equation (left) and the corresponding Dirac model (right). The purple dashed line traces a straight trajectory of the center of mass of the linear pulse. (d) Breakdown coordinate $z_{*}$ as a function of the input intensity $I_0$ estimated from the Dirac model (green curve) and 
paraxial modeling (cyan crosses). The dashed grey lines' intersection indicates the value of the input beam intensity used for (a,b,c).}
\label{fig:fig_5}
\end{figure}

As a possible implementation, we consider a realistic valley-Hall waveguide array of laser-written waveguides with parameters similar to those used in the experimental work Ref.~\cite{Noh2018}. In this case, the evolution variable $t$ becomes the longitudinal propagation distance $z$. To simulate the evolution, we solve the paraxial equation numerically in a periodic potential by propagating an optical wavepacket~\cite{Suppl}. 
For realistic laser input powers we observe a notable distortion of the beam and signatures of the catastrophe development at its trailing wavefront [Fig.~\ref{fig:fig_5}(a,b)].
Figure~\ref{fig:fig_5}(c) plots 
the intensity map in $xz$ interface plane.
The rapidly developing asymmetry 
at short distances agrees with modeling of the corresponding continuum Dirac equations and estimates of the breakdown coordinate $z_{*}$ based on Eq.~\eqref{eq:t*} [Fig.~\ref{fig:fig_5}(d)]. 

In conclusion, we have described the gradient catastrophe of the nonlinear edge wavepackets in the spinor-type Dirac equation and the formation of edge solitons at the valley-Hall domain walls. We have derived a higher-order self-steepening nonlinear Schr\"odinger equation describing these effects. Spatiotemporal numerical modeling confirmed that pulse self-steepening can manifest already in the framework of paraxial optics in weakly nonlinear media, such as topological waveguide lattices, and will likely play a key role in future experiments with topological photonic crystal waveguides. Beyond the specific valley-Hall example we considered, our findings are instructive for other emerging experimental studies of nonlinear dynamic phenomena in topological systems, such as the Chern insulators and their implementations in a variety of physical platforms spanning from metamaterials~\cite{Dobrykh2018} to optical lattices~\cite{Mukherjee2020,Xia2020} and exciton-polariton condensates~\cite{Pernet_arxiv}. 

\begin{acknowledgements}
This work was supported by the Australian Research Council (Grant DE190100430), the Russian Foundation for Basic Research (Grant 19-52-12053), the National Research Foundation, Prime Ministers Office, Singapore, the Ministry of Education, Singapore under the Research Centres of Excellence programme, and the Polisimulator project co-financed by Greece and the EU Regional Development Fund. Theoretical analysis of the continuum model was supported by the Russian Science Foundation (Grant No. 20-72-00148). D.A.S. thanks Yuri Kivshar for valuable discussions. 
\end{acknowledgements}

\newcommand{\enquote}[1]{``#1''}

\end{document}